\documentclass[twocolumn,showpacs,groupedaddress]{revtex4}
\usepackage{graphicx}

\begin{document}

\title{Quantum trajectory approach to the geometric phase: open bipartite systems}
\author{X. X. Yi,  D. P. Liu,  and  W. Wang}
\affiliation{Department of Physics, Dalian University of
Technology, Dalian 116024, China}
\date{\today}

\begin{abstract}
Through the quantum trajectory approach, we calculate the
geometric phase acquired by a bipartite system subjected to
decoherence. The subsystems that compose the bipartite system
interact with each other, and then are entangled in the evolution.
The geometric phase due to the quantum jump for both the bipartite
system and its subsystems are calculated and analyzed. As an
example, we present two coupled spin-$\frac 1 2 $ particles to
detail the calculations.
\end{abstract}

 \maketitle

Consider a quantum system that depends on some external parameters
$\vec{X}$. We are interested in the evolution of its quantum
states when the parameters $\vec{X}$ change slowly along a closed
path. For an eigenstate, such an adiabatic evolution accumulates a
geometric phase, known as the Berry phase \cite{berry84}, which
reflects the system geometry with the parameter space $\vec{X}$.
Although the  geometric phase has been rigorously formulated for
the general case of non-adiabatic, non-cyclic, and non-unitary
evolution \cite{shapere89} of  pure states,  the importance of
geometric phase in realistic systems, for instance, in the context
of geometric quantum computing
\cite{zanardi99,jones99,ekert00,falci00}, has motivated recent
interesting research into geometric phases for mixed states
\cite{uhlmann86, sjoqvist00a} and  open systems
\cite{garrison88,fonsera02,aguiera03,ellinas89,gamliel89,
kamleitner04,marzlin04,whitney03,gaitan98,nazir02,carollo03,erik04,yi05}.

Quantum trajectory (quantum jump) analyses have been applied to
certain physical systems, which show how the geometric phase for a
closed system can be modified under open system dynamics
\cite{nazir02,carollo03}. The trajectory  analyses  can be applied
for any systems evolving under a Markovian master equation or,
equivalently, under any trace preserving completely positive maps,
and  it  proves to be particularly suitable for the geometric
phase because in each particular trajectory the quantum state of
system remains pure. However, the analyses  presented in
Ref.\cite{nazir02} are only for a whole/single system, and those
in Ref. \cite{carollo03} are for conditional phase gate (it is not
a geometric phase) as well as  for a single-qubit geometric phase
gate based on numerical simulations. Then the problem of geometric
phase in coupled open bipartite  systems remains untouched.

It is more important, from the perspective of quantum computing,
to study the geometric phase in  bipartite systems, since almost
all systems employed to perform a quantum gate are composite,
i.e., it at least consists of two subsystems with direct couplings
or coupled through a third party. This together with the above
motivation stimulate the interest in study of the geometric phase
in open composite systems.

This paper contains the following interesting advances in this
field. First, we present a completely general calculation for open
bipartite systems with only one subsystem subjected to
decoherence, showing the effect of decoherence on the Berry phase
of the bipartite system. Second, having calculated the geometric
phase for the composed subsystems, we show that quantum jumps
occurring in one subsystem make  no contribution to the geometric
phase for another. Third, we identify what kind of jump operators
does not change the geometric phase for both the bipartite system
and its subsystems. Although the situation of only one subsystem
subjected to decoherence seems  less realistic, it leads us to see
how decoherence in one subsystem affects the geometric phase of
the whole system and of another subsystem, moreover the
representation for this simple situation can be easily extended to
the case of both subsystems subjected to decoherence.

We start with the most general autonomous differential equation
for the state of an open system in the Lindblad form
\cite{lindblad76}
\begin{equation}
i\hbar\frac{\partial}{\partial t}\rho=[H(\vec{X}),\rho]+{\cal
L}(\rho),\label{mseq}
\end{equation}
where $H(\vec{X})=H^{\dagger}(\vec{X})$ is the composite system
Hamiltonian depending on external parameters $\vec{X}$. ${\cal
L}(\rho)$ represents the Liouvillian, which has the general form
\cite{lindblad76}
\begin{equation}
{\cal
L}(\rho)=-\frac{i}{2}\sum_{k=1}^n\{\Gamma_k^{\dagger}\Gamma_k\rho+
\rho\Gamma_k^{\dagger}\Gamma_k-2\Gamma_k\rho\Gamma_k^{\dagger}\}.
\end{equation}
$[H(\vec{X}),\rho]$ in Eq.(\ref{mseq}) generates the coherent part
of the evolution, while ${\cal L}(\rho)$ represents the effect of
 reservoir on the dynamics of the system, the action of each
$\Gamma_k$ amounts to a different decoherence  process. Suppose
that we monitor the system and do not detect any decay, the
geometric phase for the no-jump trajectory of  the master equation
in the continuous limit is given by \cite{carollo03}
\begin{equation}
\gamma_{ab}^0=\int_0^T\frac{\langle\psi^0(t)|H|\psi^0(t)\rangle}{\langle\psi^0(t)|\psi^0(t)\rangle}dt-\mbox{arg}
\{\langle\psi^0(T)|\psi^0(0)\rangle \},
\end{equation}
where $i\frac{d}{dt}|\psi^0(t)\rangle=\tilde{H}|\psi^0(t)\rangle$,
and $\tilde{H}$ stands for the non-Hermitian effective Hamiltonian
\begin{equation}
\tilde{H}=H(\vec{X})-\frac{i}{2}\sum_{k=1}^n\Gamma_k^{\dagger}\Gamma_k.
\end{equation}
In the adiabatic limit, the geometric phase  $\gamma^0_{ab}$ for a
cyclic evolution among path $|\phi_n(\vec{X})\rangle$ yields
\cite{dattoli90}
\begin{equation}
\gamma_{n,ab}^0=\mbox{Im}\{\oint_c\langle
\Phi(\vec{X})|\nabla_{\vec{X}}\phi_n(\vec{X})\rangle d\vec{X}\},
\end{equation}
with $|\Phi_n(\vec{X})\rangle$ and $|\phi_n(\vec{X})\rangle$
satisfying
\begin{equation}
\tilde{H}|\phi_n\rangle =\lambda_n|\phi_n\rangle, \mbox{\ \ \ and
\ } \tilde{H}^{\dagger}|\Phi_n\rangle =\lambda_n^*|\Phi_n\rangle,
\end{equation}
the parameter/argument $\vec{X}$  is omitted here and hereafter
where it could not make confusion.  Now, we generalize the
formulation presented in \cite{dattoli90} for single systems to
bipartite systems. When the bipartite system undergoes an
adiabatic evolution along path $|\phi_n\rangle$, the reduced
density matrix for one subsystem, say $a$ (similarly for $b$), is
given by
\begin{equation}
\rho_a^n=Tr_b|\phi_n\rangle\langle \Phi_n|.
\end{equation}
Having written state $|\phi_n\rangle$ in the form of Schmidt
decomposition
\begin{equation}
|\phi_n\rangle=\sum_j\sqrt{p_j^n}|e_j^n\rangle_a\otimes
|f_j^n\rangle_b,
\end{equation}
the reduced density matrix $\rho_a^n$ can be expressed as
\begin{equation}
\rho_a^n=\sum_j\sqrt{p_j^n}(\sqrt{P_j^n})^* |e_j^n\rangle_a\langle
E_j^n|,
\end{equation}
where  $|E_j^n\rangle_a$ comes from the Schmidt decomposition for
$|\Phi_n\rangle$, i.e.,
\begin{equation}
|\Phi_n\rangle=\sum_j\sqrt{P_j^n}|E_j^n\rangle_a\otimes
|F_j^n\rangle_b.
\end{equation}
For the simple case where each pair of $p_i^n$ and $P_i^n$ are
$\vec{X}$-independent, the Berry phase of the bipartite system
reduces to
\begin{equation}
\gamma_{n,ab}^0=\mbox{Im}
\{\sum_j\sqrt{p_j^n}(\sqrt{P_j^n})^*(\gamma_{na,j}^0+\gamma_{nb,j}^0)\},
\end{equation}
with  $\gamma_{na,j}^0\equiv i\int _a\langle
E_j(\vec{X})|\nabla_{\vec{X}} e_j(\vec{X})\rangle_a d\vec{X}, $
and $\gamma_{nb,j}^0\equiv i\int _b\langle
F_j(\vec{X})|\nabla_{\vec{X}} f_j(\vec{X})\rangle_b d\vec{X}. $ It
is easy to see that $\gamma_{n,ab}^0$ generally is not a weighted
sum over the one particle geometric phases $\gamma_{na,j}^0$ and
$\gamma_{nb,j}^0$, since $\sqrt{p_j^n}(\sqrt{P_j^n})^*$ are
complex for open systems in general. This is different from the
Berry phase in closed bipartite system\cite{yi04}.

Now, we turn to study the effect of the quantum jump. Suppose that
the decoherence is only caused by a local reservoir, this
indicates that each $\Gamma_k\rho \Gamma_k^{\dagger}$ generates a
quantum jump  within one of the subsystem in  the trajectory.
Without loss of generality we assume here that the jumps occur
only within subsystem $a$, i.e., all $\Gamma_k$ commute with  any
operator from subsystem $b$. In the quantum trajectory analyses,
the dynamics of the bipartite system is approximated by dividing
the total evolution time $T$ into a sequence of discrete intervals
$\delta t=T/N$. The time evolution of the density matrix in each
interval takes the form \cite{carollo03} $\rho_{ab}(t+\delta
t)=\sum_{k=0}^n w_k\rho_{ab}(t)w_k^{\dagger}$, where
$w_0=1-i\tilde{H}\delta t$ and $w_k=\Gamma_k\sqrt{\delta t}$
$(k=1, ..., n)$. If there is only one jump characterized by
$\Gamma$ in the trajectory at an arbitrary time $t_1$, which
occurs in a time much shorter than any other characteristic time
of the system. Then the reduced density matrix $\rho_a$ after the
jump reads,
\begin{equation}
\rho_a^{'}(t_1)=Tr_b[ \Gamma |\psi(t_1)\rangle\langle
\psi(t_1)|\Gamma^{\dagger}]=\Gamma\rho_a(t_1)\Gamma^{\dagger},
\end{equation}
where $\rho_a(t_1)$ represents the reduced density matrix of
subsystem $a$ before the jump. Then the phase associated with the
occurrence of a jump at time $t_1$ is given by \cite{ericsson03}
\begin{eqnarray}
\gamma_a^{jump}&=&\mbox{arg}\{\sum_j p_j(t_1)\langle
\alpha_j(t_1)|\Gamma|\alpha_j(t_1)\rangle\}\nonumber\\
&=&\mbox{arg}\{\mbox{Tr}[\rho_a(t_1)\Gamma]\}.
\end{eqnarray}
Here $\rho_a(t)=\sum_j p_j|\alpha_j(t)\rangle\langle \alpha_j(t)|$
was used in the expression, this is the geometric phase of
subsystem $a$ acquired in the jump, which is obviously non-zero.
But the subsystem $b$ acquires zero geometric phase associated to
the jump, this can be understood as follows. The total phase shift
due to the jump is
\begin{equation}
\gamma_{ab}^{jump}=\mbox{arg}\{\langle\psi(t_1)|\Gamma|\psi(t_1)\rangle\},
\end{equation}
where $|\psi(t_1)\rangle$ denotes the state of the bipartite
system at the time of jump. Have writing $|\psi(t)\rangle$ into
the Schmidt decomposition,
$|\psi(t)\rangle=\sum_j\sqrt{p_j}|\beta_j(t)\rangle_a\otimes
|\tau_j(t)\rangle_b,$ we obtain
\begin{eqnarray}
\gamma_{ab}^{jump}&=&\mbox{arg}\{ \sum_j p_j  \ _a\langle
\beta_j(t)|\Gamma|\beta_j(t)\rangle_a\nonumber\\
&=&\mbox{arg}\{ \mbox{Tr} (\rho_a(t)\Gamma)\},
\end{eqnarray}
it is exactly the phase shift acquired by subsystem $a$ associated
to the jump. Thus for a bipartite system, the   subsystem that has
no change under the action of the jump operators acquires zero
geometric phase associated with the quantum jump, even if the
coupling between the subsystems are not zero. This result sharply
depends on the assumption that the jump does not need time, i.e.,
it happens immediately and lasts no time. The situation changes
when we lift the limitation/assumption on the jump, the jump in
one subsystem would transfer a geometric phase to another due to
couplings between them.

To be specific, we apply this general representation to two
coupled spin-$\frac 1 2 $ particles, in which one of the
spin-$\frac 1 2 $ particle is driven by rotating magnetic fields
and subjected to decoherence. We calculate and analyze the effect
of decoherence on the Berry phase of the bipartite system as well
as the geometric phase of the subsystems. Let us start with the
Hamiltonian that describes two coupled spin-$\frac 1 2 $ particles
in time-dependent magnetic fields,
\begin{equation}
H=\frac1 2 \alpha \vec{\sigma}_a \cdot \vec{B}(t)+
J(\sigma_a^+\sigma_b^++h.c.),
\end{equation}
where $\vec{\sigma}_j=(\sigma^x_j,\sigma^y_j, \sigma^z_j)$,
$\sigma^i_j$ are the pauli operators  for subsystem $ j (j=a,b)$
and $\sigma_j^+=(1/2)(\sigma_j^x+i\sigma_j^y).$ We will choose
$\vec{B}(t)=B_0 \hat{n}(t)$ with the unit vector
$\hat{n}=(\sin\theta\cos\phi,\sin\theta\sin\phi,\cos\theta)$ and
have assumed  that only the subsystem $a$ is driven by the
external field. The classical field $\vec{B}(t)$ acts as an
external control parameter, as its direction and magnitude can be
experimentally altered. $J$ stands for the constant of coupling
between the two spin-$\frac 1 2 $ particles.  This coupling is not
a typical spin-spin coupling, but rather a toy model describing a
double spin flip; nevertheless,  the presentation in this paper
may be generalized to the system of nuclear magnetic
resonance(NMR) \cite{jones99}, where we can use Carbon-13 labelled
chloroform in $d_6$ acetone as the sample, in which the single
$^{13}C$ nucleus and the $^1H$ nucleus play the role of the two
spin-$\frac 1 2 $ particles. The  constant of spin-spin coupling
$J\sigma_1^z\sigma_2^z$  in this case is  $J\simeq (2\pi)214.5
\mbox{Hz}$, and we may control the rescaled coupling constant
$g=2J/\alpha B_0$ by changing the magnitude of the external
magnetic field. We would like to address that the interaction
between the two spin-$\frac 1 2 $ particles in our model is not a
typical spin-spin coupling as that in NMR. So, we have to make a
mapping when we employ the presentation in NMR system and when all
subsystem are driven by  classical fields. The Liouvillian which
describes the decoherence effect in subsystem $a$ may have the
form ${\cal L}(\rho)=-\frac{i}{2} \{\sigma_a^+\sigma_a^-\rho+
\rho\sigma_a^{+}\sigma_a^--2\sigma_a^-\rho\sigma_a^{+}\}.$ The
corresponding  non-Hermitian Hamiltonian reads
$\tilde{H}=H-\frac{1}{2}\kappa \sigma_a^+\sigma_a^-,$ dissipation
would  give rise to modify the eigenvalues and the corresponding
eigenvectors that are given by
\begin{eqnarray}
|\Psi_j\rangle&=&\frac{1}{\sqrt{M_j}}[a_j(\phi,\theta,g)|eg\rangle+
b_j(\phi,\theta,g)|ee\rangle\nonumber\\
&+&c_j(\phi,\theta,g)|gg\rangle+d_j(\phi,\theta, g)|ge\rangle],
\label{eigenv1}
\end{eqnarray}
with
\begin{eqnarray}
a_j(\phi,\theta,g)&=& \sin\theta e^{-i\phi},
c_j(\phi,\theta,g)=E_j-\cos\theta-\frac{i}{2}\kappa,\nonumber\\
d_j(\phi,\theta,g)&=&\frac{g(\cos\theta-E_j-\frac i 2 \kappa
)\sin\theta}{\sin^2\theta-(\cos\theta-\frac{i}{2}\kappa)^2+E_j^2}
e^{i\phi},\nonumber\\
b_j(\phi, \theta,
g)&=&-\frac{\cos\theta+E_j-\frac{i}{2}\kappa}{\sin\theta}e^{-i\phi}
d_j(\phi,\theta,g),\nonumber\\
M_j&=&|a_j|^2+|b_j|^2+|c_j|^2+|d_j|^2,\label{eigenf}
\end{eqnarray}
and
\begin{widetext}
\begin{eqnarray}
E_1&=&\sqrt{\sin^2\theta+(\cos\theta-\frac{i}{2}\kappa)^2+\frac
{g^2}{ 2} +\frac g 2 \sqrt{g^2+4\sin^2\theta}}=-E_2,
\nonumber\\
E_3&=&\sqrt{\sin^2\theta+(\cos\theta-\frac{i}{2}\kappa)^2+\frac
{g^2}{2} -\frac g 2 \sqrt{g^2+4\sin^2\theta}}=-E_4. \label{eigenv}
\end{eqnarray}
\end{widetext}
The eigenvectors and the corresponding eigenvalues of
$(\tilde{H})^{\dagger}$ have the same form as those of $\tilde{H}$
in Eq.(\ref{eigenf},\ref{eigenv}), but  $(-i\kappa)$ should be
replaced by $i\kappa$. We will use  notations of $A_j, B_j, C_j$,
and $D_j$ $(j=1,2,3,4)$ corresponding to $a_j,b_j, c_j$, and $d_j$
in Eq. (\ref{eigenf}) as the coefficients that appear in the
instantaneous eigenstates of $\tilde{H}^{\dagger}$. In fact,
$A_j(i\kappa)=a_j(-i\kappa)$, $B_j(i\kappa)=b_j(-i\kappa)$ and so
on.
\begin{figure}
\includegraphics*[width=0.8\columnwidth,
height=0.6\columnwidth]{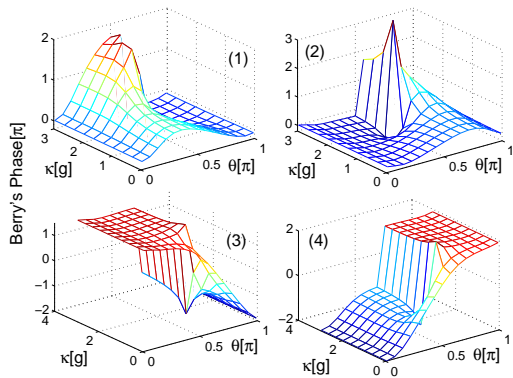} \caption{ The Berry phase
$\gamma_{n,ab}^0$ for the bipartite system in a no-jump trajectory
as a function of the decoherence rate $\kappa$ and the azimuthal
angle $\theta$. (1)-(4) in the figure correspond to different
instantaneous eigenstates $|\Psi_j\rangle$, $j=1,2,3,4$. }
\label{fig1}
\end{figure}
The numerical results for the Berry phase with no-jump were
presented in figure 1, where we plot $\gamma_{n,ab}^0$
$(n=1,2,3,4)$ as a function of the spontaneous rate $\kappa$ and
the azimuthal $\theta$. In contrast with the case of $\kappa=0$
\cite{yi04}, there are jumps at $\theta=\pi/2$ with
$\kappa>\kappa_0$ depending on the path the system follows. The
jumps appearing in figure 1 may be understood as follows. The
evolution of the state along the no-jump trajectory represented by
one of equation (\ref{eigenv1}) can be mapped on the Bloch sphere
with spontaneous decay rate $\sim \kappa$. The evolution is then a
smooth spiral converging to the lower state, thus the geometric
phase increases due to the spontaneous decay when the initial
state falls onto the upper semi-sphere; while the phase decrease
when it initially is in the lower semi-sphere. Therefore, the
geometric phase has a jump at the crossover point $\theta=\pi/2$.
This was schematically shown in figure 2.
\begin{figure}
\includegraphics*[width=0.4\columnwidth,
height=0.4\columnwidth]{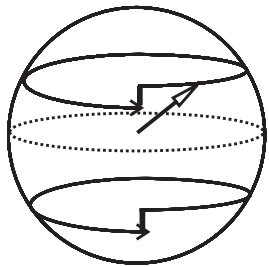} \caption{ A schematic
illustration for the  effect of spontaneous decay.  The solid
angle increases due to the decay when the state is on the upper
sphere, while it decreases  for that on the lower sphere. }
\label{fig1}
\end{figure}
In the case where only one quantum jump (described by operator
$\sigma_a^-$) occurs at any time $t_1$, the geometric phase shift
due  to the jump is given by
\begin{equation}
\gamma_{n,ab}^{jump}=\mbox{arg}\{a_nC_n^* +b_nD_n^* \}, \mbox{\ \
}(n=1,2,3,4),
\end{equation}
selected numerical results for $\gamma_{n,ab}^{jump}$ were
illustrated in figure 3. The quantum jump occurring at a time of
$\phi=\pi$ was assumed for this plot. Clearly, singularity appears
at $\theta=\pi/2$, where the subsystem $a$ experiences a crossover
from the upper half of the Bloch sphere to the lower one.

\begin{figure}
\includegraphics*[width=0.8\columnwidth,
height=0.6\columnwidth]{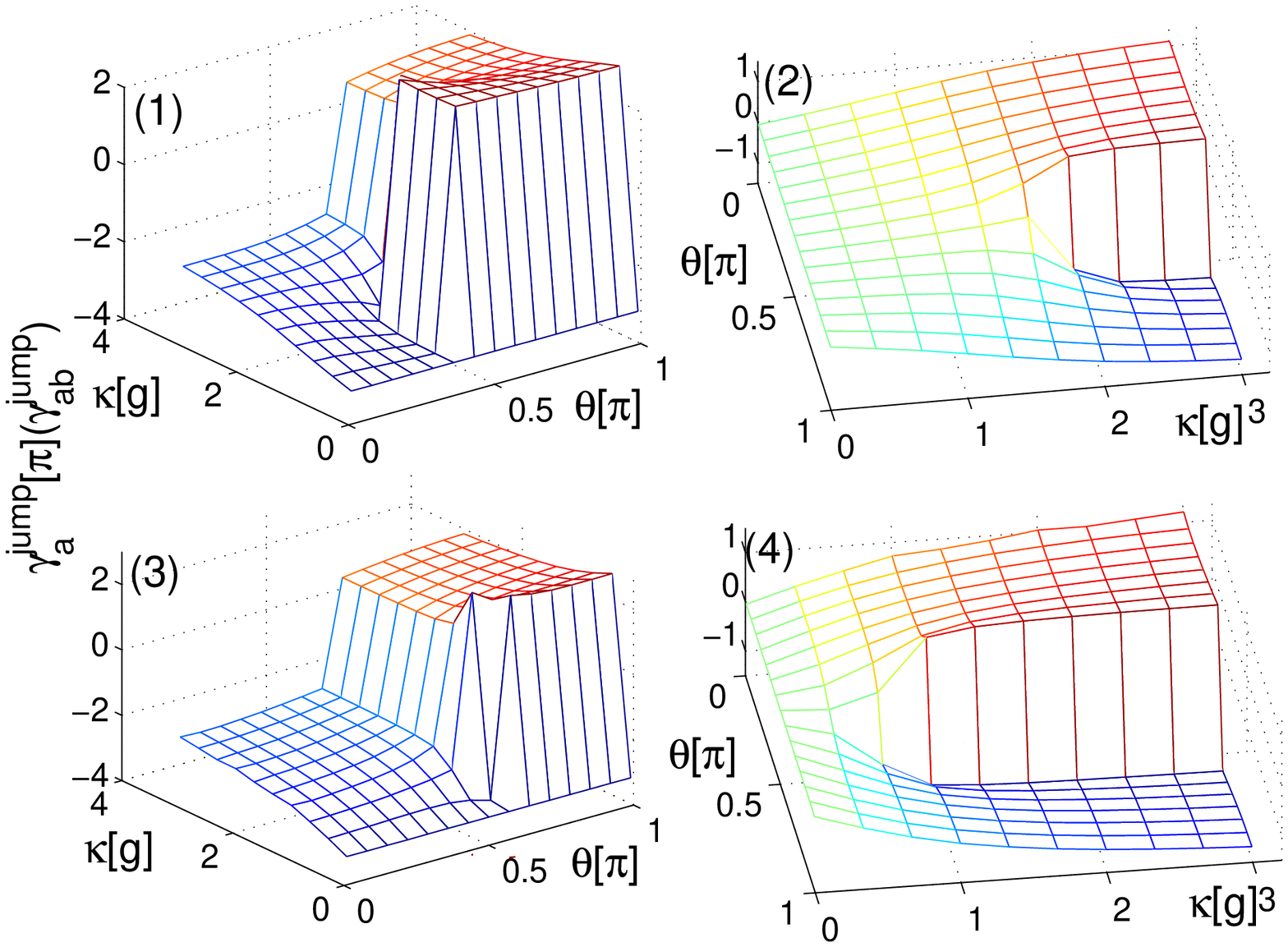} \caption{Geometric phase
acquired by the bipartite system (or subsystem $a$) in a single
quantum jump among the adiabatic path $|\Psi_j\rangle$. The
quantum jump was assumed to occur at the time when $\phi=\pi$. }
\label{fig2}
\end{figure}

In conclusion, we have investigated the geometric phase in open
bipartite systems. This study is of relevance to the geometric
quantum computing, where geometric phase may be used to perform
quantum information processing. To our best knowledge, this issue
remains unaddressed by the trajectory approach, in particular for
coupled open systems.  The results show that there is a
singularity in the dependance of the geometric phase on the
azimuthal angle with a specific $\kappa>\kappa_0$, where
$\kappa_0$ depends on the inter-subsystem coupling constant $g$.
The phase shift due to the quantum jump was also calculated, it
has shown that the phase shift depends strongly on the direction
to which the spin-$\frac 1 2 $ particle points. The jump occurs at
$\theta=\pi/2$ can be interpreted in picture of the Bloch sphere,
in which the decaying of the subsystem  $a$ results in a smooth
spiral converging to the ground state, so when the state falls in
the upper semi-sphere, the decay increases the geometric phase,
but it lowers the geometric phase when the state on another
semi-sphere, this leads to the jump in the phase at
$\theta=\pi/2$. And it is interesting to note that there is no
jump when the decay rate $\kappa < \kappa_0$, the critical value
of the decay, this is due to that the system finishes a cyclic
evolution before the system decay.

\ \ \\
 This work was
supported by NCET of M.O.E, and NSF of China Project No. 10305002.\\

\end{document}